\documentclass[a4paper,11pt]{article}
\usepackage{pos}
\usepackage{subfig}
\renewcommand{\logo}{\relax}

\usepackage{lineno}
\title{Simulated performance and calibration of CMS Phase-2 Upgrade Inner Tracker sensors}
\ShortTitle{Simulated performance and calibration of CMS Phase-2 IT sensors}

\author*[a]{Tamas Almos Vami}
\author[a]{Morris Swartz}
\onbehalf{on behalf of the CMS Collaboration}
\affiliation[a]{Johns Hopkins University,\\
  Baltimore, MD, USA}

\emailAdd{Tamas.Almos.Vami@cern.ch}
\emailAdd{morris@jhu.edu}

\abstract{The next upgrade of the Large Hadron Collider (LHC) is planned from 2026 when the collider will move to its High Luminosity phase (HL-LHC). The CMS detector needs to be substantially upgraded during this period to exploit the fourfold increase in luminosity provided by the HL-LHC. This upgrade is referred to as the CMS Phase-2 Upgrade. A program of laboratory and beam test measurements, and performance studies based on the detailed simulation of the detector was carried out to support the decision of the technology of the sensors to be adopted in the different regions of the detector for the Phase-2 Upgrade. Among the various options considered, CMS chose to use 3D sensors with a 25\,$\times$\,100 $\mu$m$^2$ pixel cell in the innermost layer of the barrel and planar sensors with a 25\,$\times$\,100 $\mu$m$^2$ pixel cell elsewhere. In this paper, we detail the simulation studies that were carried out to choose the best sensor design. These studies include a detailed standalone simulation of the sensors made with PixelAV and the expected performance on high level observables obtained with the simulation and reconstruction software of the CMS experiment.}

\FullConference{% 10th International Workshop on Semiconductor Pixel Detectors for Particles and Imaging (Pixel2022),\\
  12-16 December 2022\\
  Santa Fe, New Mexico, USA}

\begin{document}

\newcommand{\mumS}{$\mu \text{m}^2\,$}

\newcommand{\neutroneq}{$\text{n}_{\text{eq}}/\text{cm}^2$}

\maketitle

\section{Introduction}

%Figure~\ref{fig:LHCschedule} details the schedule of the Large Hadron Collider (LHC). It shows the shutdown periods (long shutdown, LS) and the data taking periods.
%In the more detailed schedule in the bottom, red corresponds to shutdown periods, while green corresponds to data taking periods. 
The Large Hadron Collider (LHC) will be upgraded to its High Luminosity phase (HL-LHC) in the shutdown period from 2026. The HL-LHC will reach the instantaneous luminosity of $7.5\,\times\,10^{34} \,\,\,\text{s}^{-1}\text{cm}^{-2}$ which is four times the Run-2 value of $7.5\,\times\,10^{34} \,\,\,\text{s}^{-1}\text{cm}^{-2}$. The number of simultaneous proton-proton interactions per 25 ns bunch crossing (pileup) is expected to be between 140 and 200, which is a similar fourfold increase with respect to the current value of 55. The current detectors would not be able to operate in such conditions, which motivates the need to upgrade them. This is referred to as the Phase-2 Upgrade of CMS.

%\begin{figure}[h!]
%    \centering
%    \includegraphics[width=.92\textwidth]{Figures/LHCschedule.png}
%    \caption{Schedule of the LHC showing when the HL-LHC starts its operation.}
%    \label{fig:LHCschedule}
%\end{figure}

\begin{figure}[h!]
    \centering
    \includegraphics[width=.72\textwidth]{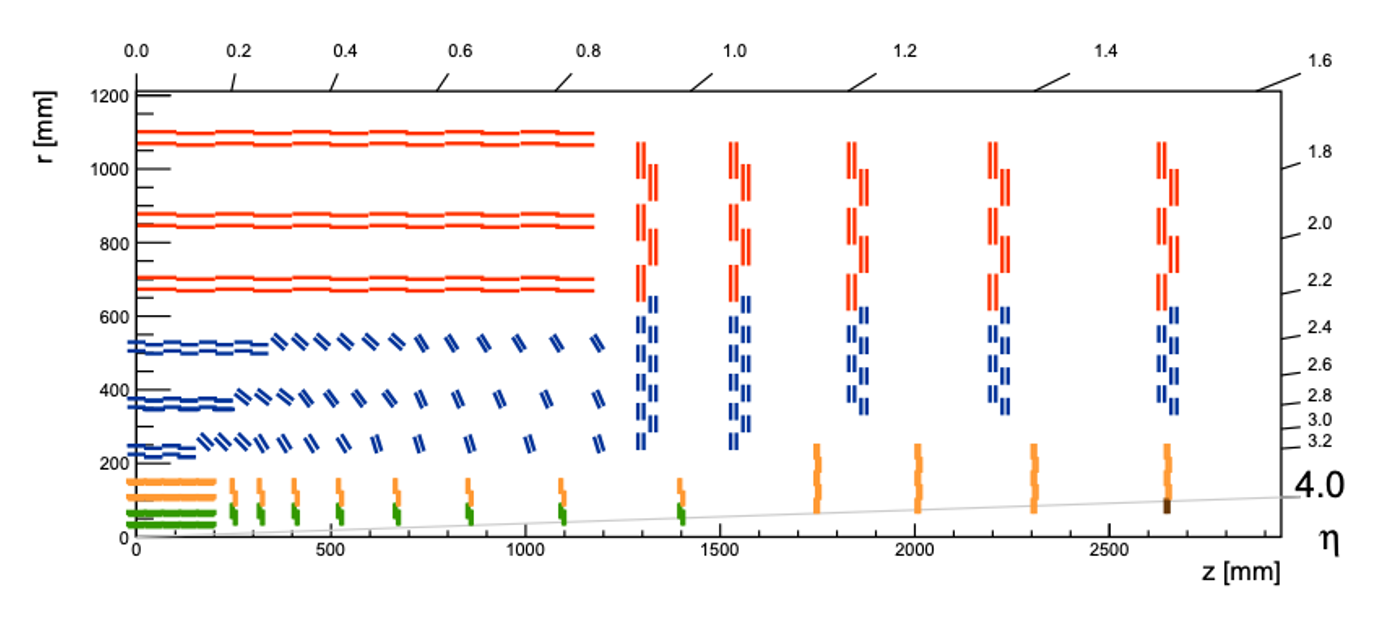}
    \caption{Sketch of one quarter of the Phase-2 CMS tracker.  Outer Tracker modules are represented in red and blue, while Inner Tracker modules are represented in green and yellow.}
    \label{fig:Phase2Tracker}
\end{figure}

The upgraded CMS tracker detector \cite{CERN-LHCC-2017-009} will provide  increased acceptance for tracking up to $|\eta| < 4$ and significantly reduced mass through the use of carbon fiber mechanics, and $\text{CO}_2$ cooling. The layout of the upgraded tracker  is shown in Fig.~\ref{fig:Phase2Tracker}. It consists of two parts, the Outer Tracker (OT) and the Inner Tracker (IT). The OT has  six barrel layers and five pairs of forward disks, while the IT has four barrel layers and twelve pairs of forward disks. 

The IT has 3900 hybrid modules which use either two or four read out chips per module. The read out chips are designed by the RD53 collaboration \cite{Chistiansen:1553467}. Sensors are bump bonded to the readout chip in a 50\,$\times$\,50\,\mumS pattern. The active thickness of the sensors is 150\,$\mu$m for all the scenarios considered. This paper details the sensors properties used in the IT, such as the pixel dimensions and technologies used.

%\newpage
Simulation studies presented in this paper were performed with both the standard CMS software environment (CMSSW, \cite{CMSSW}) and PixelAV \cite{Chiochia:2004qh}. PixelAV is independent from CMSSW, but can perform more detailed simulations. 

%The following topics are discussed in this paper.
%\begin{itemize}
%    \item PixelAV simulation description
%    \item Sensor size studies (50\,$\times$\,50\,\mumS versus 25\,$\times$\,100\,\mumS)
%    \item Sensor technology studies (planar versus 3D)
%    \item Simulation of avalanche gain effect in planar sensors
 %   \item Tracking performance and heavy flavor tagging performance
%\end{itemize}

%\newpage
\section{PixelAV simulation description}

PixelAV simulations use a charge deposition model based on the Bichsel pion-Si cross sections~\cite{RevModPhys.60.663}. 
%The cross section as a function of energy is shown in Fig.~\ref{fig:BichselAndTCAD}.
The delta-ray range is calculated using the continuously slowing-down approach with NIST ESTAR dE/dx data~\cite{1992estar}. 
The carrier transport on a given charge (q) is based on the Runge-Kutta integration of the saturated drift velocity (\textbf{v}). The velocity is calculated as shown  in Equation~\ref{eq:satDriftVel} for a given electric (\textbf{E}) and magnetic (\textbf{B}) field, where $\mu$ is the mobility and $ r_H$ is the Hall factor.

\begin{equation}
    \textbf{v} = \frac{\mu [q\textbf{E} + \mu r_H \textbf{E} \times \textbf{B} + q \mu^2 r_H^2 (\textbf{E} \cdot \textbf{B})\textbf{B}]}{1 + \mu^2 r_H^2 |\textbf{B}|^2}
    \label{eq:satDriftVel}
\end{equation}

The electric field is derived from an ISE TCAD simulation of a pixel cell.
%, as shown in Fig.~\ref{fig:BichselAndTCAD}
%, where we observe a non-uniform E-field.
The TCAD simulations include effects for charge trapping, diffusion, and induction on implants. After this, there is an electronics simulation, where we apply noise, linearity, thresholds, and miscalibration effects.

%\begin{figure}[h!]
%   \centering
%       \subfloat{
%       \centering
%       \includegraphics[width=.42\textwidth]{Figures/BichselXsections.png}
%   }
%    \subfloat{
%       \centering
%       \includegraphics[width=.42\textwidth]{Figures/TCADexample.png}
%   }
%   \caption{Bichsel cross sections (left, \cite{RevModPhys.60.663}) and TCAD electric field example (right).}
%   \label{fig:BichselAndTCAD}
%\end{figure}

Irradiation simulations for the Phase-2 sensors are based on models developed for the 2018 Phase-1 detector (based on data corresponding to a fluence of $1\times 10^{15}$\,\neutroneq), but are scaled to the fluence expected from the HL-LHC. The read-out chip threshold is 1000 electrons for all cases discussed below.

Cross talk with neighboring pixels depends on the geometry; pixel size of 25\,$\times$\,100\,\mumS\,  generates 10\% crosstalk, while the  50\,$\times$\,50\,\mumS generates no crosstalk. The bias voltage assumptions for the 25\,$\times$\,100\,\mumS sensors start with 350 V, and for 50\,$\times$\,50\,\mumS start with 100 V. In both cases, the bias voltage (HV) is increased to to maximize the signal efficiency.

Simulations are evaluated by comparing detector resolution as a function of track angle ($\eta$). We use the same calibrations and reconstruction algorithm as used in CMSSW. 
Resolution is obtained by taking the RMS of the residual distribution between measured and expected hit positions, which better accounts for non-Gaussian tails. These measurements are  performed in two charge bins: 0 < $Q/Q_{avg}$ < 1  and 1 < $Q/Q_{avg}$ < 1.5, where $Q$ is the charge and $Q_{avg}$ is the average charge produced by the hit.

Another important parameter is the charge collection efficiency (CCE) which is defined as the ratio of collected charge to all charge generated.

\section{Pixel size studies}

\begin{figure}[h!]
    \centering
    \includegraphics[width=.99\textwidth]{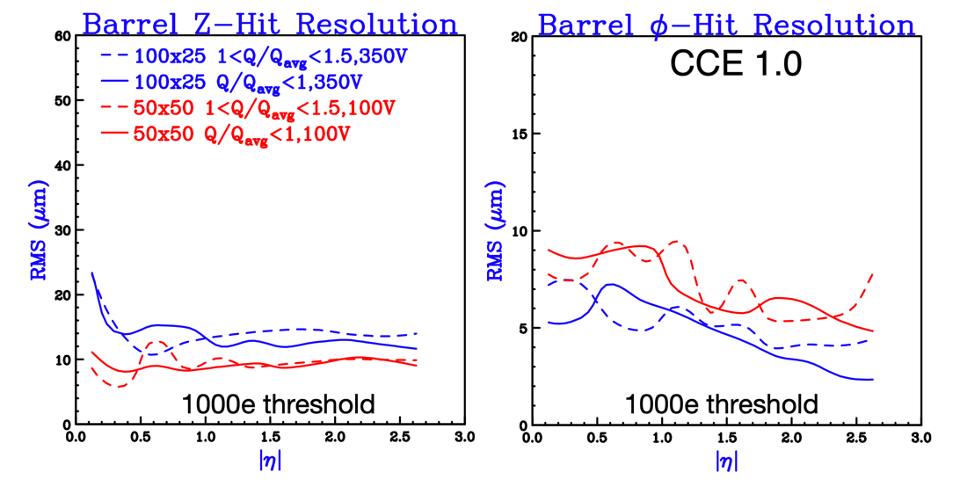}
    \caption{The resolution (RMS) as a function of the track angle for unirradiated sensors, along the beam direction (left) and in the azimuth direction (right). The solid line is for the charge bin 0 < $Q/Q_{avg}$ < 1  and the dashed line is for the charge bin 1 < $Q/Q_{avg}$ < 1.5.}
    \label{fig:SizeChoiceUnirrad}
\end{figure}

In this section, we compare the pixel size options of 50\,$\times$\,50\,\mumS and 25\,$\times$\,100\,\mumS. Figure~\ref{fig:SizeChoiceUnirrad} shows the resolution (RMS) as a function of the track angle for unirradiated sensors in both the $z$ coordinate (along the beam), and the $\phi$ coordinate (azimuthal angle). Red lines correspond to the  50\,$\times$\,50\,\mumS choice, while the blue lines correspond to the  25\,$\times$\,100\,\mumS choice.  The CCE for the unirradiated case is 100\%.

We observe that the 50\,$\times$\,50\,\mumS performs better in the beam direction, while the 25\,$\times$\,100\,\mumS\,  performs better in the azimuth direction. The local maxima of the curve is at boundaries when transitioning from a single pixel cluster to a double pixel cluster, or higher cluster sizes.

\begin{figure}[h!]
    \centering
    \includegraphics[width=.99\textwidth]{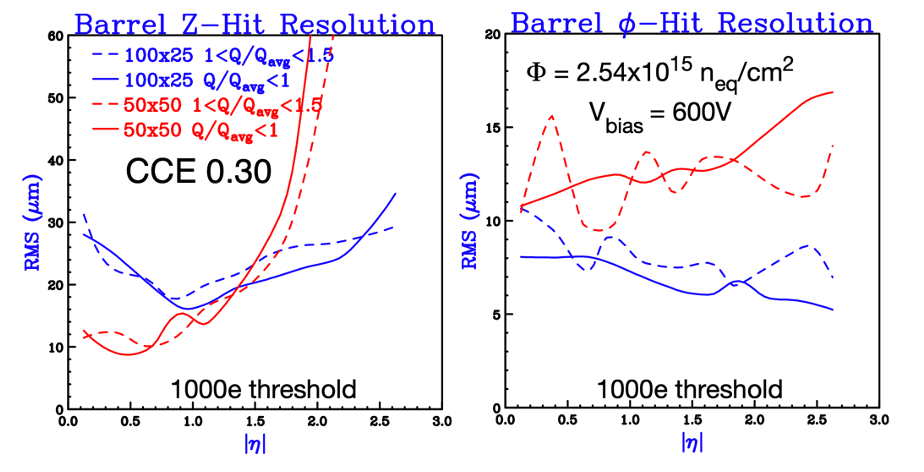}
    \caption{The resolution (RMS) as a function of the track angle for irradiated sensors, along the beam direction (left) and in the azimuth direction (right).}
    \label{fig:SizeChoiceIrrad}
\end{figure}

Figure~\ref{fig:SizeChoiceIrrad} shows the resolutions but for irradiated sensors after 3000 fb$^{-1}$, which corresponds to the end of HL-LHC running and a fluence of $2.53 \times 10^{15}$\,\neutroneq. In this case the bias voltage is increased to 600 V to maximise the signal efficiency.

The situation is very different from the unirradiated case. Now the 50\,$\times$\,50\,\mumS case only performs better in the low $\eta$ regions, but breaks down at high $\eta$. On the other hand, the 25\,$\times$\,100\,\mumS still performs better in the azimuth direction. Based on these results the 25\,$\times$\,100\,\mumS is the better choice in terms of the pixel dimensions.

\newpage
\section{Sensor technology studies}

\begin{figure}[h!]
    \centering
    \includegraphics[width=.72\textwidth]{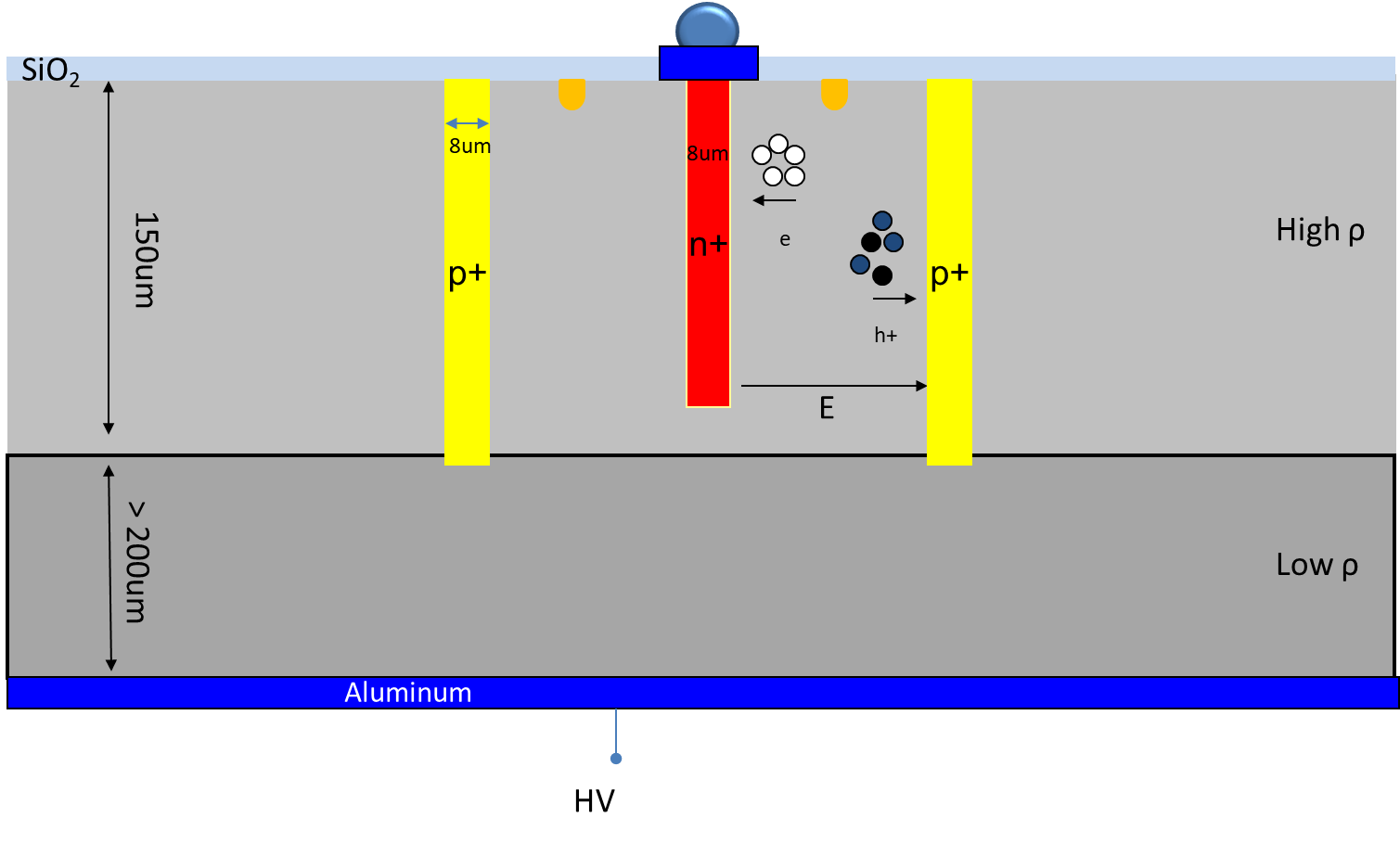}
    \caption{Sketch of a 3D sensor. Electrons are collected on the n+ implant (red column) while holes are collected in the p+ implants (yellow columns).}
    \label{fig:3DsensorSketch}
\end{figure}

Besides planar sensors, the possibility of 3D sensors was considered. Figure~\ref{fig:3DsensorSketch} shows how 3D sensors differ from planar sensors by collecting charge on columnar implants that penetrate the substrate.
%This creates a radial electric field and the path length of carriers is reduced. Electrons are collected on the n+ implant while holes are collected in the p+ implants.

\newpage
The original version of PixelAV used a segmented parallel plate capacitor model to estimate the signal induced by trapped carriers. While this works well for planar sensors, it cannot be used to describe more complex, less symmetric implant geometries.  PixelAV was therefore modified to use a lookup table based implementation of the Ramo-Shockley weighting potential. This is a general method that works for 3D sensors as well.

\begin{figure}[h!]
    \centering
    \includegraphics[width=.32\textwidth]{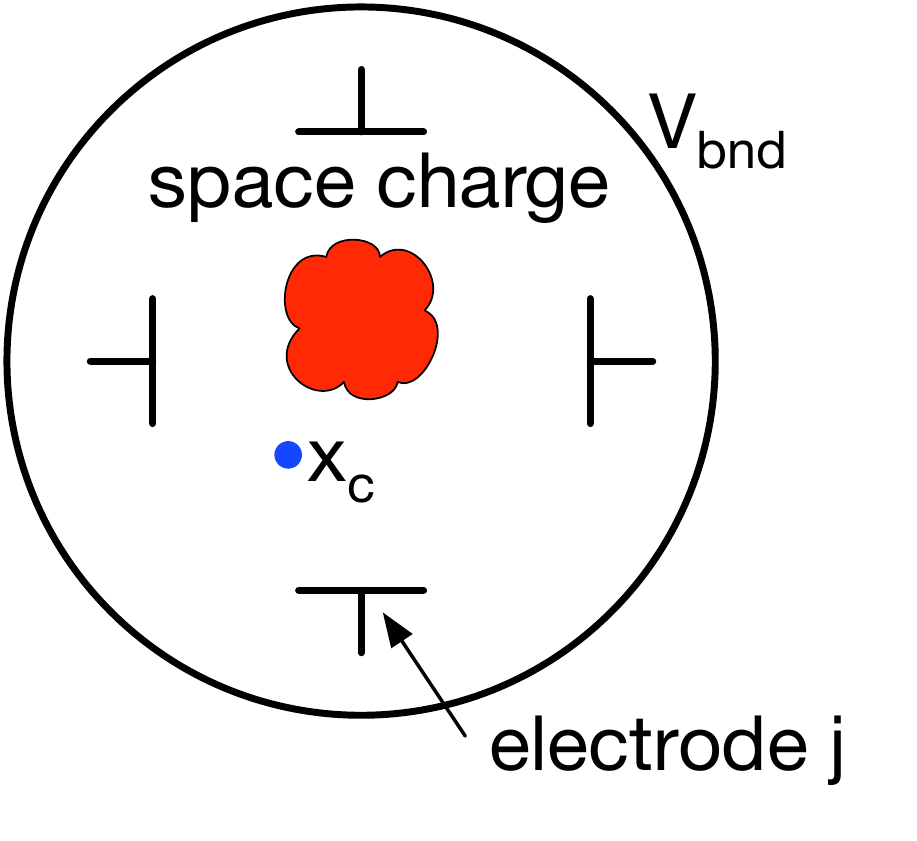}
    \caption{Sketch of electrode system for 3D sensors.}
    \label{fig:ramo_thm}
\end{figure}

The potential function $\phi(x)$ is the solution of Laplace's equation for a system of electrodes with $V_j = V_0$, $V_{bnd}=0$, and $V_i = 0$ for $i \neq j$ as shown in Fig.~\ref{fig:ramo_thm}.

Charge on electrode $j$ ($Q_j$) induced by carrier at $x_c$ is 
\begin{equation}
    Q_j = \frac{q_c \phi(x_c)}{V_0}  
\end{equation}

\noindent
where the $\frac{\phi(x_c)}{V_0}$ is the weighting potential and $q_c$ is the charge of the carrier.

One further complication for 3D sensors is that TCAD 9.0 does not support meshing across region boundaries to calculate the weighting potential. To resolve this, we place equipotential conducting "contacts" on the inside surfaces of square voids to represent the implants in a 2.5 x 2.5 pixel array.

\begin{figure}[h!]
    \centering
    \includegraphics[width=.72\textwidth]{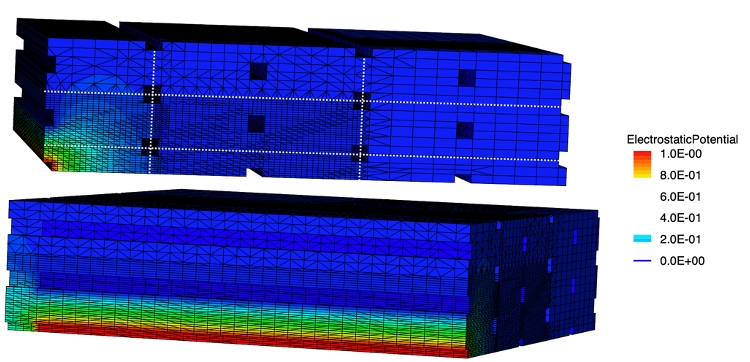}
    \caption{Electric field maps generated by TCAD for the 3D sensors.}
    \label{fig:TCAD_3D}
\end{figure}

After these modifications, the electric field maps generated by TCAD for the 3D sensors can be seen in Fig.~\ref{fig:TCAD_3D}.

\begin{figure}[h!]
    \centering
    \includegraphics[width=.99\textwidth]{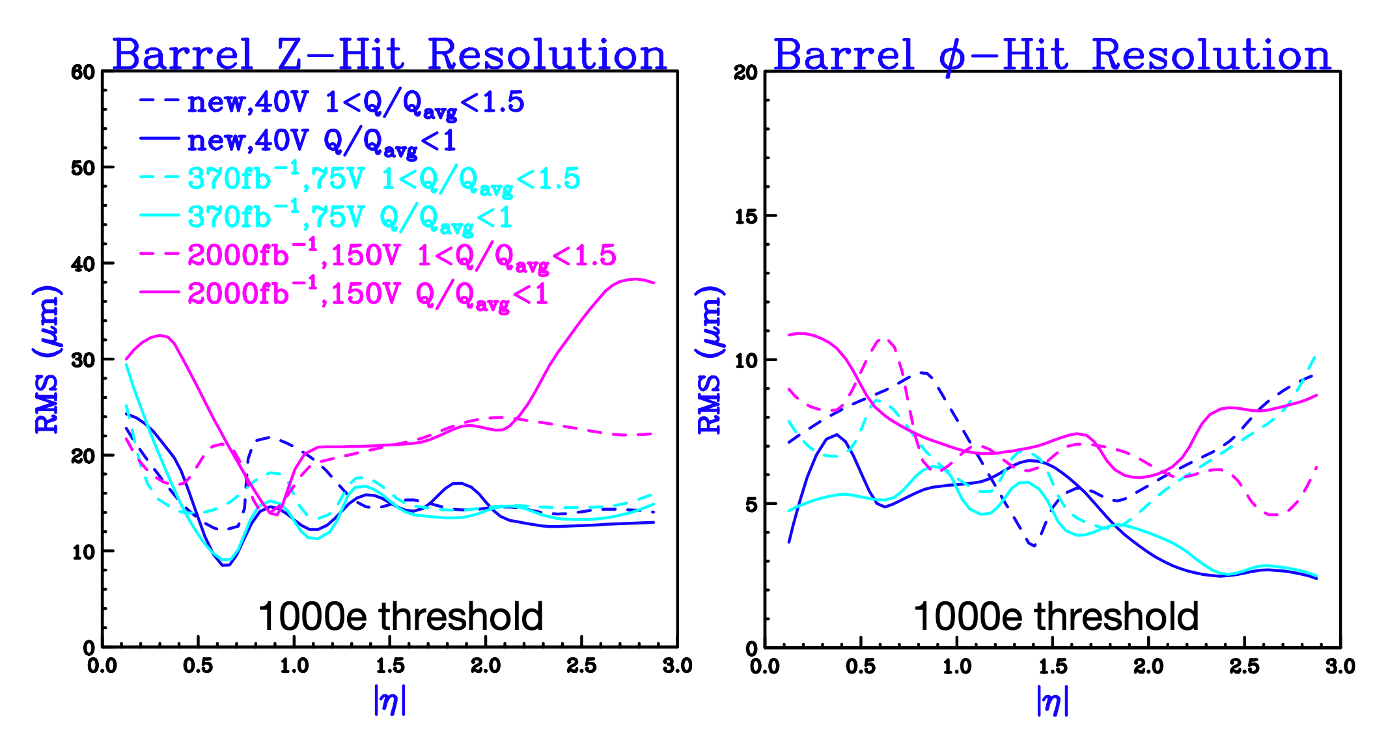}
    \caption{The residuals as a function of the track angle for the 3D sensor case for three irradiation scenarios; dark blue corresponds to new sensors, turquoise corresponds to irradiation after 370 fb$^{-1}$, and magenta is for irradiation after 2000 fb$^{-1}$, along the beam direction (left) and in the azimuth direction (right).}
    \label{fig:TechnoStudies}
\end{figure}

Figure~\ref{fig:TechnoStudies} shows the residuals as a function of the track angle for the 3D sensor case for three irradiation scenarios; dark blue corresponds to new sensors operated at 40 V, turquoise corresponds to sensors operated at 75 V, after 370 fb$^{-1}$ irradiation, and magenta is for 150 V with irradiation after 2000 fb$^{-1}$.

We can observe that the 3D sensors after 370 fb$^{-1}$ perform similarly to unirradiated sensors, while at 2000 fb$^{-1}$ the resolutions show the effect of charge loss. The CCE is reduced to 84\% for irradiation after 370 fb$^{-1}$  and to 32\% for irradiation after 2000 fb$^{-1}$.

\newpage
Table \ref{tab:summaryScencario3D} shows the summary of different irradiation scenarios for the 3D sensors and their performance. The sensors after irradiation at 2000 fb$^{-1}$ need 150 V bias voltage (HV) setting, otherwise significant cluster breakage is observed. Resolutions in the beam and azimuth directions are averaged over all pseudorapidity.

\begin{table}[h!]
    \centering
     \begin{tabular}{|c || c | c | c|} 
     \hline
     Scenario & New & 370 fb$^{-1}$ & 2000 fb$^{-1}$ \\ [0.5ex] 
     \hline
     \hline
     Fluence & 0 \neutroneq & $3\times 10^{15}$\,\neutroneq  &  $17 \times10^{15}$\,\neutroneq \\ 
     \hline
     Bias & 40 V & 75 V & 150 V\\
     \hline
     Resolution (Z) & 13.9\,$\mu$m & 14.3\,$\mu$m & 22.5 \,$\mu$m\\
     \hline
     Resolution ($\phi$) & 5.6\,$\mu$m & 5.9\,$\mu$m & 9.8\,$\mu$m\\
     \hline
     CCE & 0.96 & 0.84  & 0.39\\
     \hline
     \end{tabular}
     \caption{Summary of different irradiation scenarios for the 3D sensors and their performance.}
     \label{tab:summaryScencario3D}
\end{table}

In summary, 3D sensors have excellent performance after irradiation with resolutions comparable to current detector performance. Based on these plots 3D sensors are a better choice for the first layer of the IT.

\section{Simulation of avalanche gain effect}

The avalanche gain effect is non-negligible for high bias voltage values for the Phase-2 planar sensors. This section details how to include this effect in the PixelAV simulations using test beam data from DESY. The test beam data corresponds to an irradiation of $4\times10^{15}$\,\neutroneq\, (denoted as "Data 40x") and contains several bias voltage settings. We studied the charge profiles, i.e. the path length normalized pixel charge as a function of the sensor depth, using each setting. We observe that lower bias voltage settings (less than 600\,V) are better described by the simulations, while for the higher bias voltage settings, the peak charge and the total charge both deviate from the predictions of the simulation.

\begin{figure}[h!]
   \centering
   \subfloat{
       \centering
       \includegraphics[width=.42\textwidth]{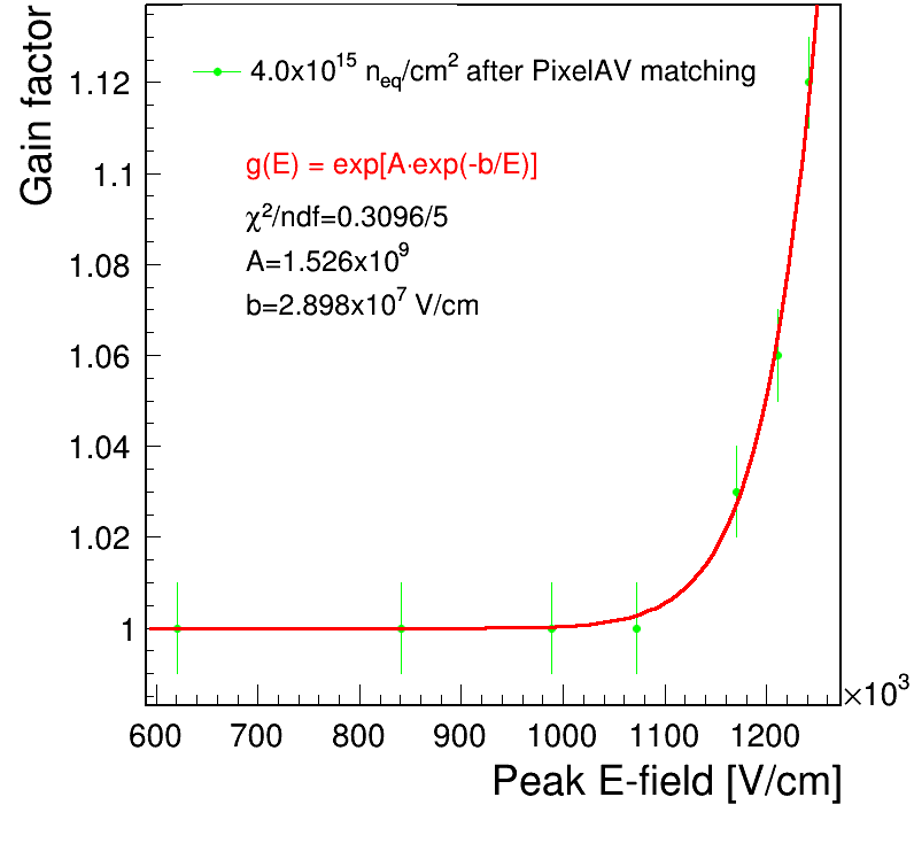}
   }
    \subfloat{
       \centering
       \includegraphics[width=.42\textwidth]{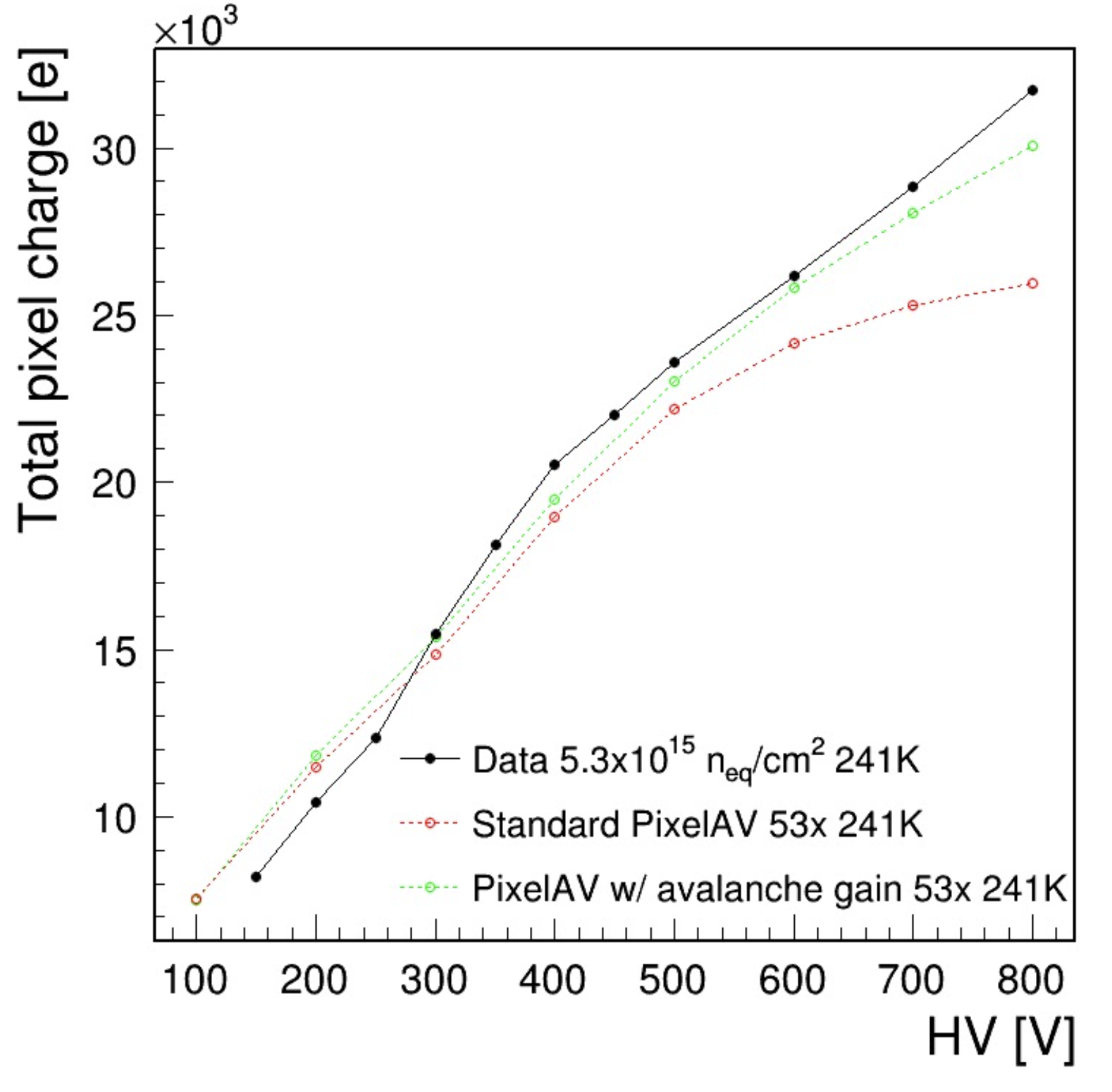}
   }
    \caption{Selected gain factors as a function of the peak electric field extracted from TCAD (left). The validation of the method (right) using the total charge as a function of the bias voltage setting for an orthogonal set of test beam data.}
    \label{fig:GainFVsPeakEAndSummeryGainAvalanche}
\end{figure}

In order to improve on this,  PixelAV was updated to include a gain factor when collecting electrons.
%The induced charge from trapped carriers would not experience any gain. When testing the code with gain = 1, we get identical results to vanilla pixelAV. 
We performed a scan for each sample with a gain factor variation 
%No other change is include, all of these are at 253K and the same irradiation assumption. The resulting charge profiles for selected bias voltage settings (800 V in the right, 700 V in the middle, and 600 V in the right) are shown in Fig.~\ref{fig:GainFactorScan}. Black points correspond to the test beam data, while different colors are PixelAV simulations with different gain factors.
%\begin{figure}[h!]
%    \centering
%    \includegraphics[width=.99\textwidth]{Figures/GainFactorScan.png}
%    \caption{Charge profiles for selected bias voltage settings (800 V in the right, 700 V in the middle, and 600 V in the right) after the gain scan. Black points correspond to the test beam data, while different colors are PixelAV simulations with different gain factors.}
%    \label{fig:GainFactorScan}
%\end{figure}
 and selected the gain factor that describes the charge profile of the data the best.
 %For $\text{HV} < 600\,\text{V}$, PixelAV the plots are not shown as the gain=1.0 describes the data sufficiently. 
Figure~\ref{fig:GainFVsPeakEAndSummeryGainAvalanche} shows the selected gain factors as a function of the peak electric field extracted from TCAD.
The uncertainty of the gain factor value is estimated to be 0.01. Reference \cite{Serezhkin} suggests the function g(E):
\begin{equation}
    g(E) = e^{A\cdot\text{exp}\left(-\frac{b}{E}\right)}
\end{equation}
where $A\cdot\text{exp}\left(-\frac{b}{E}\right)$ is the coefficient of the impact ionization for electrons/holes, and $b$ is the parameter for the breakdown electric field.

To validate this procedure, we used test beam data with sensors irradiated at $5.3\times10^{15}$\,\neutroneq\, and compared against simulations in which we included the avalanche gain effect with the parameterization from the  $4 \times 10^{15}$\,\neutroneq\, data studies. The results are shown in Fig.~\ref{fig:GainFVsPeakEAndSummeryGainAvalanche}. Black points correspond to the test beam data, the red points are the standard PixelAV simulation and the green points are PixelAV with the avalanche gain effect. For the PixelAV simulations (red and green) we rescaled the fluences to  $5.3 \times 10^{15}$\,\neutroneq. 

The charges in the simulation without avalanche gain effect deviate from the data at higher bias voltages, while the simulation including avalanche gain effect is better at describing the test beam data.

\section{CMSSW full simulations}

\begin{figure}[h!]
   \centering
    \subfloat{
       \centering
       \includegraphics[width=.92\textwidth]{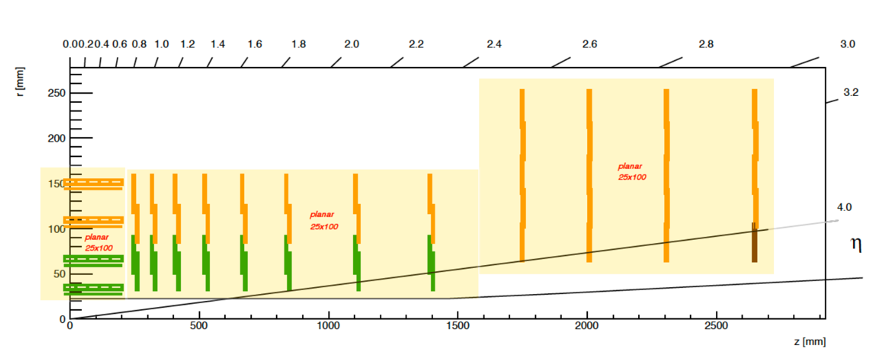}
   }\\
   \vspace{-2em}
    \subfloat{
       \centering
       \includegraphics[width=.92\textwidth]{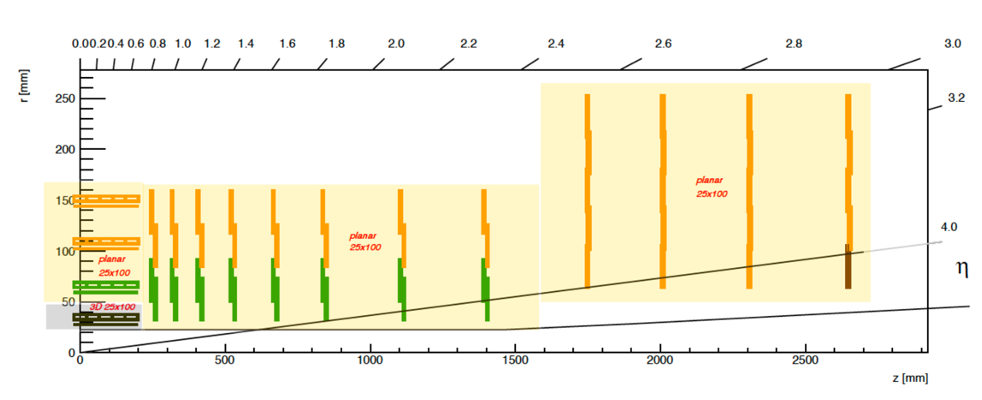}
   }\\
   \vspace{-2em}
    \subfloat{
       \centering
       \includegraphics[width=.92\textwidth]{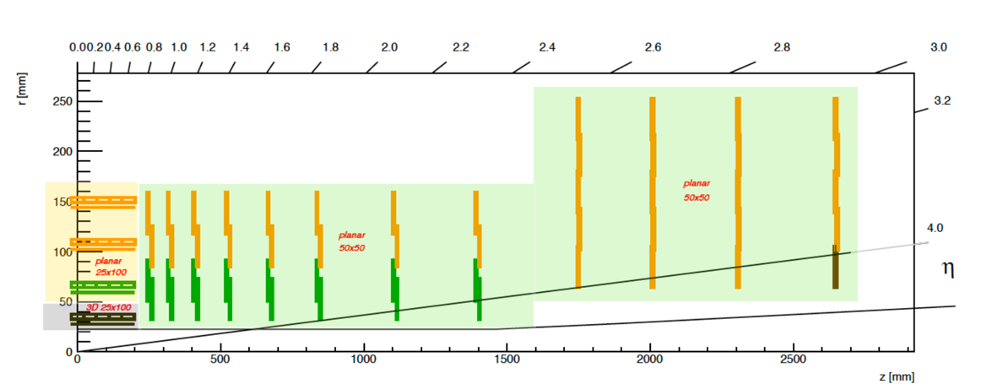}
   }
   \caption{Three IT layouts that are being compared in this section:  25\,$\times$\,100\,\mumS in the both the barrel and the endcaps (T21, top), 25\,$\times$\,100\,\mumS in the both the barrel and the endcaps, but 3D sensors in the first barrel layer (T25, middle), 25\,$\times$\,100\,\mumS in the barrel, with 3D sensors in the first barrel layer, and 50\,$\times$\,50\,\mumS planar sensors the endcaps (T25, bottom).}
   \label{fig:phase2ITLayouts}
\end{figure}

This section compares the scenarios detailed in the above sections using CMSSW. It should be noted that there is no radiation simulation yet in CMSSW, so all these results correspond to unirradiated sensors. Physics performance was evaluated for three layouts of the IT, which are shown in Fig.~\ref{fig:phase2ITLayouts}. These layouts sketch a quarter of the Inner Tracker in the r-z view generated using the \texttt{tkLayout} tool \cite{tkLayout}. Modules with 1x2 readout chips are shown in green, modules with 2x2 readout chips are  shown in orange. The three layouts are defined as:
\begin{itemize}
    \item T21: 25\,$\times$\,100\,\mumS in the both the barrel and the endcaps.
    \vspace{-.5em}
    \item T25: 25\,$\times$\,100\,\mumS in the both the barrel and the endcaps, but 3D sensors in the first barrel layer.
    \vspace{-.5em}
    \item T26: 25\,$\times$\,100\,\mumS in the barrel, with 3D sensors in the first barrel layer, and 50\,$\times$\,50\,\mumS planar sensors the endcaps.
\end{itemize}

\noindent
We used two different samples of 80 000 simulated events at $\sqrt{s}=14 \,\text{TeV}$:
\begin{itemize}
    \item Single-track (muon) events for tracking performance studies, consisting of a flat $p_T$ spectrum of [1, 200] GeV, without smearing of the primary vertex.
    \vspace{-.5em}
    \item $t\bar{t}$ events with pileup of 200 for heavy flavor tagging studies, generated with an approximately 4 cm Gaussian smearing of the beamspot along the beamline.
\end{itemize}

\begin{figure}[h!]
    \centering
    \includegraphics[width=.72\textwidth]{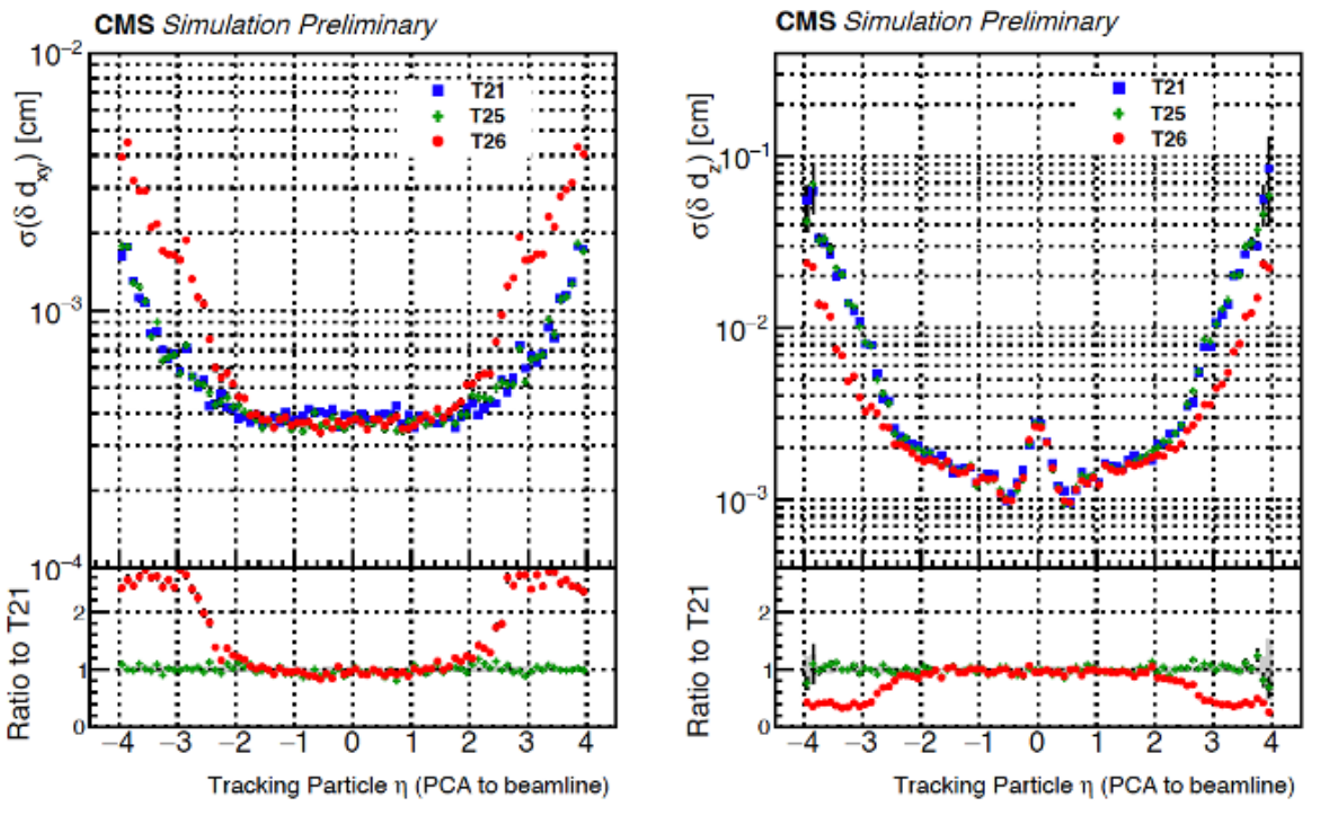}
    \caption{Resolution on the track impact parameter in the transverse (left) and longitudinal (right) directions, as a function of track angle ($\eta$) for different geometries. Blue points correspond to the T21 geometry, green to the T25 geometry and the red to the T26 geometry.}
    \label{fig:TrackingPerf}
\end{figure}

In terms of tracking performance, the resolution on the track impact parameter as a function of track angle is shown in Fig.~\ref{fig:TrackingPerf}. Blue points correspond to the T21 geometry, green to the T25 geometry and the red to the T26 geometry. Resolution is computed for tracks which have at least one-hit in the IT. The result from T21 layout is used as the reference (denominator) in the ratio plots.

\newpage
The comparison of 25\,$\times$\,100\,\mumS planar (T21) and 25\,$\times$\,100\,\mumS 3D sensors (T25 and T26) for barrel-only tracks, $|\eta|<1.2$, shows that the resolution on track impact parameter, both in the transverse dimension ($d_{xy}$) and the longitudinal dimension ($d_{z}$), is the same when the first barrel layer is built with planar or 3D sensors.
%This is ascribed to the large coverage in $\phi$ of modules in layer 1, approximately [-16$^{\circ}$, +16$^{\circ}$], which is larger than the typical Lorentz angle before irradiation (11$^{\circ}$). The large $\phi$ acceptance guarantees charge sharing in 3D sensors despite the absence of Lorentz drift. On the contrary, the coverage in $\phi$ of modules in layer 2 is smaller, approximately [-8$^{\circ}$, +8$^{\circ}$], which disfavors the usage of 3D sensors in this layer. 
Comparison for endcap-only tracks, $|\eta|>2.7$, shows that the improvement of the resolution in one projection is approximately of the same amount as the deterioration in the other.

\begin{figure}[h!]
    \centering
    \includegraphics[width=.99\textwidth]{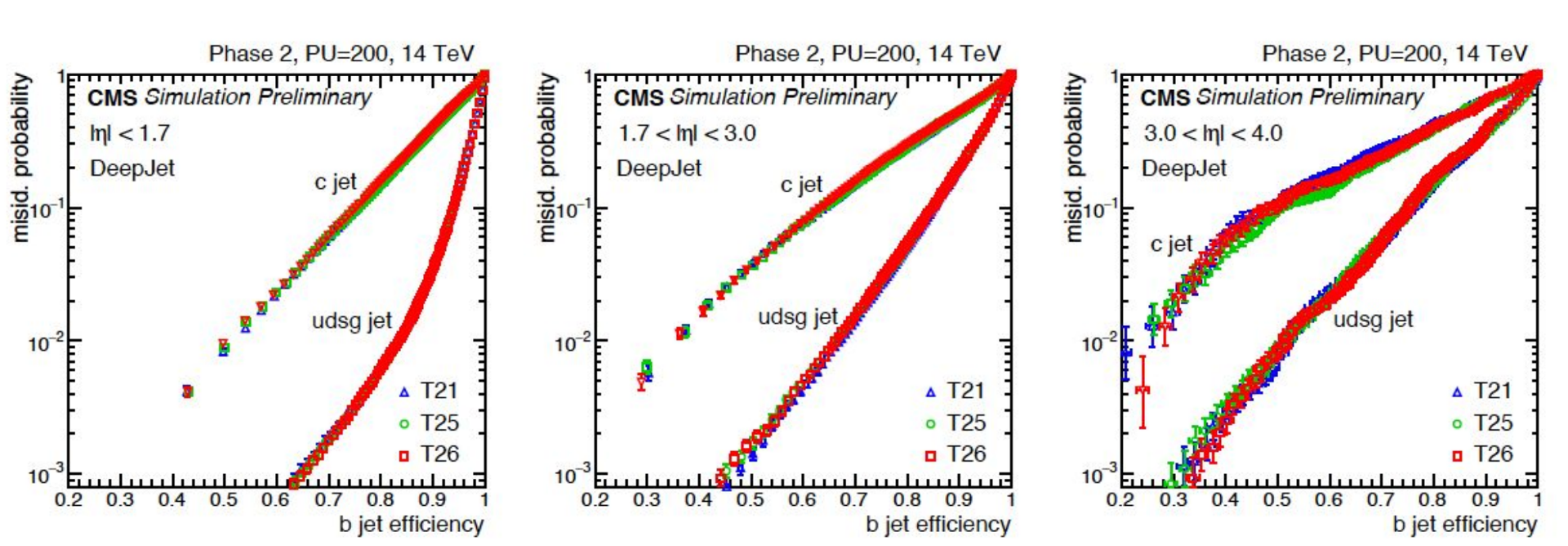}
    \caption{The ROC curves for light-/charm-jet contamination as a function of b-jet efficiency for the “DeepJet” tagger \cite{Bols_2020} for the layouts T21 (blue), T25 (green) and T26 (red) in three regions of $\eta$: $|\eta| < 1.7$, $1.7 < |\eta| < 3.0$, $3.0 < |\eta| < 4.0$}.
    \label{fig:bTagPerformance}
\end{figure}

Heavy flavor tagging performance is shown in Fig.~\ref{fig:bTagPerformance}, where ROC curves describe light-/charm-jet contamination as a function of b-jet efficiency for the “DeepJet” tagger \cite{Bols_2020} in three regions of $\eta$: $|\eta| < 1.7$, $1.7 < |\eta| < 3.0$, $3.0 < |\eta| < 4.0$. The edges of the bins were defined to minimize migration effects. The corresponding number of jets per bin is 360k , 105k and 15k, respectively. 

For central jets (|$\eta$|\,<\,3.0), Fig.~\ref{fig:bTagPerformance} shows that there is no difference in performance between the three geometries.  This result is expected as the simulated impact parameter resolution for single muons is identical for each layout.  For forward jets (|$\eta$|\,>\,3.0) however, the performance differential between T26 and T21/T25 geometries is minor and suggests that pixel sizes of 25\,$\times$\,100\,\mumS could be used throughout the Inner Tracker.

These finding are in line with the simulations of the irradiated sensors discussed in the earlier chapters. With all these considerations, the final design choice is T25, i.e. 3D sensors in the first layer, and planar sensors everywhere else, with a pixel size of 25\,$\times$\,100\,\mumS.

\newpage
\section{Conclusions}

In this paper, the HL-LHC and CMS Phase-2 Inner Tracker project was introduced. PixelAV, TCAD simulations were detailed, especially concentrating on how they are used to simulate irradiated sensors. Different pixel sizes (50\,$\times$\,50\,\mumS and 25x100\,\mumS) and sensor technologies (planar and 3D sensors) were studied. The development in PixelAV to simulate 3D sensors using the Ramo--Shockley theorem and weighting potentials was presented. Simulations to DESY test beam data were compared and the avalanche gain effect for planar sensors was characterized. The tracking and heavy flavor tagging performance of three layouts in CMSSW were studied. Following these and other studies, the decision has been made to use 3D sensors in the first layer of the barrel, and use planar sensors in everywhere else. The decision for the size of the pixels is to use 25x100\,\mumS. 

\bibliography{main}

\end{document}